\documentclass{aastex} 
\usepackage{spr-astr-addons}
\usepackage{hyperref}
\usepackage{url}
\usepackage{lscape}

\shorttitle{A new class of solutions}
\shortauthors{Pandya \textit{et al}}

\begin{document}

\title{Modified Finch and Skea stellar model compatible with observational data}

\author{D. M. Pandya} 
\affil{Department of Mathematics and Computer Science, Pandit Deendayal Petroleum University, Raisan, Gandhinagar 382 007, India}
\email{dishantpandya777@gmail.com}
 
\author{V. O. Thomas}
\affil{Department of Mathematics, Faculty of Science, The Maharaja Sayajirao University of Baroda, Vadodara 390 001, India}
\email{votmsu@gmail.com}

\author{R. Sharma}
\affil{Department of Physics, P. D. Women's College, Jalpaiguri 735 101, India}
\email{rsharma@iucaa.ernet.in}

%
%

\begin{abstract}

We present a new class of solutions to the Einstein's field equations corresponding to a static spherically symmetric anisotropic system by generalizing the ansatz of Finch and Skea [{\em Class. Quantum Grav.} \textbf{6} (1989) 467] for the gravitational potential $g_{rr}$. The anisotropic stellar model previously studied by Sharma and Ratanpal (2013) [\emph{Int. J. Mod. Phy. D} \textbf{13} (2013) 1350074] is a sub-class of the solutions provided here. Based on physical requirements, regularity conditions and stability, we prescribe bounds on the model parameters. By systematically fixing values of the model parameters within the prescribed bound, we demonstrate that our model is compatible with the observed masses and radii of a wide variety of compact stars like 4U 1820-30, PSR J1903+327, 4U 1608-52, Vela X-1, PSR J1614-2230, SAX J1808.4-3658 and Her X-1.   

\end{abstract}


\section{\label{sec1}Introduction}
\cite{Finch89}, making use of \cite{Duorah87} ansatz for the metric potential $g_{rr}$ corresponding to a static spherically symmetric perfect fluid space-time, developed a stellar model which was later shown to comply with all the physical requirements of a realistic star by \cite{Delgaty98}. Consequently, the Finch-Skea model has been explored by many investigators in different astrophysical contexts, particularly for the studies of cold compact stellar objects (see for example, \cite{Hansraj06,Tik09,Ayan13}). One noticeable feature of the \cite{Finch89} model is that it assumes isotropy in pressure. However, theoretical investigations of \cite{Ruderman72} and \cite{Canuto74}, amongst others have shown that anisotropy might develop in the high density regime of compact stellar objects. In other words, radial and transverse pressures might not be equal at the interior of ultra-compact stars. \cite{Bowers74} have extensively discussed the conditions under which anisotropy might occur at stellar interiors which include presence of type-3A super fluid, electromagnetic field, rotation etc. They have also
established the non-negligible effects of local anisotropy on the maximum equilibrium mass and surface redshift of the distribution. Accordingly, different anisotropic stellar models have been developed and effects of anisotropy on physical properties of stellar configurations have been analyzed by many investigators, viz. \citet{Maharaj89,Gokhroo94,Patel95,Tikekar98,Tikekar99,Tikekar05,Thomas05,Thomas07}. Impacts of anisotropy on the stability of a stellar configuration have been studied by \citet{Dev02,Dev03,Dev04}. \cite{Sharma07} and \cite{Thirukkanesh08} have obtained analytic solutions of compact anisotropic stars by assuming a linear equation of state(EOS). To solve the Einstein-Maxwell system, \cite{Komathiraj07} have used a linear equation of state. By assuming a linear EOS, \cite{Sunzu14} have reported solutions  for a charged anisotropic quark star. \cite{Feroze11} and \cite{Maharaj12} have used a quadratic-type EOS for obtaining solutions of anisotropic distributions. \cite{Varela10} have analyzed charged anisotropic configurations admitting a linear as well as non-linear equations of state. For a star composed of quark matter in the MIT bag model, \cite{Paul11} have shown how anisotropy could effect the value of the Bag constant. For a specific polytropic index, exact solutions to Einstein's field equations for an anisotropic sphere admitting a polytropic EOS have been obtained by \cite{Thirukkanesh12}. \cite{Maharaj13b} have used the same type of EOS to develop an analytical model describing a charged anisotropic sphere. Polytropes have also been studied by \\ 
\cite{Nilsson01}, \cite{Heinzle03} and \cite{Kinasiewicz07}. \cite{Thirukkanesh14} have used modified Van der Waals EOS to represent anisotropic charged compact spheres. For specific forms of the gravitational potential and electric field intensity, \cite{Malaver14} has  prescribed solutions for a stellar configuration whose matter content admits a quadratic EOS. \cite{Malaver113} and \cite{Malaver213} have also found exact solutions to the Einstein-Maxwell system using the Van der Waals modified EOS. 

Recently, \cite{Sharma13}, making use of the \cite{Finch89} ansatz, have generated a class of solutions describing the interior of a static spherically symmetric anisotropic star. In this paper, we have generalized the \cite{Sharma13} model by incorporating a dimensionless parameter $n (> 0)$ in the \cite{Finch89} ansatz and assumed the system to be anisotropic, in general. We have shown that such assumptions can provide physically viable solutions which can be used to model realistic stars. Implications of the modified ansatz (by including an adjustable parameter $n$) on the size and physical properties of resultant stellar configurations have been analyzed. Based on physical requirement, we have put constraints on our model parameters and subsequently shown that a wide variety of observed pulsars can be accommodated within the prescribed bound of the model parameters. In particular, we have shown that the predicted masses and radii of pulsars like 4U 1820-30, PSR J1903+327, 4U 1608-52, Vela X-1, PSR J1614-2230, SAX J1808.4-3658 and Her X-1 can well be achieved by systematically fixing the parameter $n$. Most importantly, for a given mass, it is possible to constrain the radius so as to get the desired compactness by fixing the \emph{compactness parameter} $n$ in this model. 

The paper has been organized follows: In \ref{sec2}, for a static spherically symmetric anisotropic fluid sphere, we have solved the relevant field equations by making a particular choice of the metric potential $g_{rr}$ which is a generalization of the \cite{Finch89} ansatz. In \ref{sec3}, we have laid down the boundary conditions and in \ref{sec4}, we have put constraints on the model parameters based on physical requirements, regularity conditions and stability. Physical applications of our model have been discussed in \ref{sec5}. In \ref{sec6}, we have concluded by pointing out the main results of our model.

\section{\label{sec2} Modified Finch and Skea model}  
We write the interior space-time of a static spherically symmetric distribution of anisotropic matter in the form
\begin{equation}
ds^2 = e^{\nu (r)} dt^2 -  e^{\lambda(r)}dr^2 - r^2 (d\theta^2 + sin^2 \theta d\phi^2),
\label{d}
\end{equation}
where, 
\begin{equation}
e^{\lambda} = \left(1 + \frac{r^2}{R^2}\right)^{n}.\label{ansz}
\end{equation}
In (\ref{ansz}), $n > 0$ is a dimensionless parameter and $R$ is the curvature parameter having dimension of a length. Note that the ansatz (\ref{ansz}) is a generalization of the \cite{Finch89} model which can be regained by setting $n = 1$.  

We follow the treatment of \citet{Maharaj89} and write the energy-momentum tensor of the anisotropic matter filling the interior of the star in the form
\begin{equation}
T_{ij} = \left(\rho + p \right) u_i u_j - p g_{ij} + \pi_{ij},
\label{f}
\end{equation}
where, $ \rho$ and $p$ denote the energy-density and isotropic pressure of the fluid, respectively and $ u_i $ is the $4$-velocity of the fluid. The anisotropic stress-tensor $\pi_{ij}$ has the form
\begin{equation}
\pi_{ij} = \sqrt{3}S \left[C_iC_j - \frac{1}{3}(u_iu_j - g_{ij})\right],
\label{g}
\end{equation}
where, $ C^i = (0,-e^{-\lambda/2},0,0) $. For a spherically symmetric anisotropic distribution, $S(r)$ denotes the magnitude of the anisotropic stress. The non-vanishing components of the energy-momentum tensor are the following:
\begin{equation}
T_0^0 = \rho, ~~~ T_1^1 = - \left(p + \frac{2S}{\sqrt{3}}\right), ~~~ T_2^2 = T_3^3  =  -\left(p - \frac{S}{\sqrt{3}}\right).
\label{h}
\end{equation}
Consequently, radial and tangential pressures of the fluid can be obtained as
\begin{eqnarray}
p_r &=& -T_1^1 = \left(p + \frac{2S}{\sqrt{3}}\right),\label{i}\\
p_\perp &=& -T_2^2 = \left(p - \frac{S}{\sqrt{3}}\right),\label{j}
\end{eqnarray}
so that 
\begin{equation}
S = \frac{p_r - p_\perp}{\sqrt{3}}.
\label{k}
\end{equation}
The potentials of the space-time metric (\ref{d}) and physical variables of the distribution are related through the Einstein's field equations
\begin{eqnarray}
8\pi \rho &=& \frac{1 - e^{-\lambda}}{r^2} + \frac{e^{-\lambda} {\lambda}^\prime}{r},\label{l}\\
8\pi p_r &=& \frac{e^{-\lambda} - 1}{r^2} - \frac{e^{-\lambda}{\nu}^\prime}{r},\label{m}\\
8\pi p_\perp &=& e^{-\lambda} \left[\frac{\nu^{\prime\prime}}{2} + \frac{{\nu^\prime}^2}{4} - \frac{\nu^\prime \lambda^\prime}{4} + \frac{\nu^\prime - \lambda^\prime}{2r}\right].\label{n}
\end{eqnarray}

\noindent By defining the mass $ m(r) $ within a radius $ r $ as
\begin{equation}
m(r) = 4\pi \int\limits_{0}^{r} u^2 \rho(u) du,
\label{o}
\end{equation}
we get an equivalent description of the system as
\begin{eqnarray}
e^{-\lambda} &=& 1 - \frac{2m}{r},\label{p}\\
r (r - 2m) \nu^{\prime} &=& 8\pi p_r r^3 + 2m,\label{q}\\
-\frac{4}{r} (8\pi \sqrt{3}S) &=& (8\pi\rho + 8\pi p_r)\nu^{\prime} + 2 (8\pi p_r^{\prime}).\label{r}
\end{eqnarray}
Using (\ref{ansz}) in (\ref{l}) and (\ref{p}), we obtain the energy-density and mass $ m(r) $ in the form
\begin{eqnarray}
8\pi \rho &=& \frac{\frac{1}{r^2}{\left(1 + \frac{r^2}{R^2}\right) \left[\left(1 + \frac{r^2}{R^2}\right)^n-1\right]}+\frac{2 n}{R^2}}{\left(1 + \frac{r^2}{R^2}\right)^{n+1}},\label{s}\\
m(r) &=& \frac{\frac{r}{2}\left[\left(1 + \frac{r^2}{R^2}\right)^n-1\right]}{\left(1 + \frac{r^2}{R^2}\right)^n}.
\label{t}
\end{eqnarray}
To integrate Eq.~(\ref{q}), following \cite{Sharma13}, we write the radial pressure in the form
\begin{equation}
8\pi p_r = \frac{p_0 \left(1-\frac{r^2}{R^2}\right)}{R^2 \left(1 + \frac{r^2}{R^2}\right)^{n+1}},
\label{u}
\end{equation}
which is a reasonable assumption since the radial pressure vanishes at $ r = R$. Consequently, the curvature parameter $R$ in our model turns out to be the boundary of the star. Substituting (\ref{u}) in (\ref{q}) and integrating, we get
\begin{equation}
e^{\nu} = C \left(1 + \frac{r^2}{R^2}\right)^{p_0} exp{\left[-\frac{p_0 r^2}{2 R^2} + \int\limits_{0}^{r}\left[\left(1 + \frac{u^2}{R^2}\right)^{n} - 1 \right] \frac{1}{u} du\right]},
\label{v}
\end{equation}
where, $ C $ is a constant of integration. 

Finally, using Eqs.~(\ref{r}), (\ref{s}) and (\ref{u}), the anisotropy is obtained as 
\begin{equation}
8\pi \sqrt{3}S = A_1 (r) - \left\{A_2 (r) (A_3 (r) + A_4 (r) )\right\},
\label{x}
\end{equation}
where, 
\[A_1 (r) = \frac{p_0 \frac{r^2}{R^2}\left[(n+2)-\frac{n r^2}{R^2}\right]}{R^2 \left(1 + \frac{r^2}{R^2}\right)^{n+2}},\] 
\[A_2 (r) = \frac{1}{4} \left[\left(1 + \frac{r^2}{R^2} \right)^{n-1}\right ]-\frac{p_0}{4} \frac{r^2}{R^2} + \frac{p_0}{2} \frac{\frac{r^2}{R^2}} {\left(1 + \frac{r^2}{R^2}\right)}, \]
\[A_3 (r) = \frac{\left(1 + \frac{r^2}{R^2}\right) \left[\left(1 + \frac{r^2}{R^2}\right)^n-1\right]\frac{1}{r^2}+ \frac{2 n}{R^2}}{\left(1 + \frac{r^2}{R^2}\right)^{n+1}},\]
\[A_4 (r) = \frac{p_0 \left(1-\frac{r^2}{R^2}\right)}{R^2 \left(1 + \frac{r^2}{R^2}\right)^{n+1}}.\]
Note that the anisotropy vanishes at the center $ r = 0 $, as expected. Subsequently, the tangential pressure can be obtained from the relation
\begin{equation}
8\pi p_{\perp} = 8\pi p_r - 8\pi \sqrt{3} S.
\label{y}
\end{equation}
Using the above relations, we also obtain
\begin{eqnarray}
\nonumber
\frac{d p_r}{d \rho} &=& \frac{p_0 \frac{r^4}{R^4}[(n + 2) - n \frac{r^2}{R^2}]}{(1 + \frac{r^2}{R^2})^{n+2} - \left[1 + \left\{(n + 2) + (1 - n - 2 n^2)\frac{r^2}{R^2}\right\} \frac{r^2}{R^2}\right]},\label{hh}\\
\nonumber
\frac{d p_\perp}{d \rho} &=& \frac{1}{c^2}\frac{d p_r}{d \rho} - \frac{p_0 \frac{r^4}{R^4} \left[I(r) + D(r)\right] + R^{6} B(r)}{4R^{6}\left(1 + \frac{r^2}{R^2}\right)^{n+3} E(r)},\label{ii}
\end{eqnarray}
where,
\[B(r) = F(r) + G(r) + H(r),\]
\[F(r) = \left[\left\{1 + \frac{r^2}{R^2}\left(1 - n - 2n^2\right)\right\} \left(1 + \frac{r^2}{R^2}\right)\frac{r^2}{R^2}\right],\]
\[G(r) = \left[1 - (n-1)\frac{r^2}{R^2}\right]\left(1 + \frac{r^2}{R^2}\right)^{2n+2},\]
\[H(r) = -2\left(1 + \frac{r^2}{R^2}\right)^{n+1} \left[1 + 2 \frac{r^2}{R^2} - (n - 1) \frac{r^4}{R^4}\right],\]
\begin{eqnarray}
\nonumber
I(r) = 2\left[R^6\left\{\left(1 - \frac{3r^2}{R^2}\right)-\left(7 - \frac{r^2}{R^2}\right)\frac{nr^2}{R^2} \right.\right. \\ \nonumber
\left.\left. + 2 \left(1 + \frac{r^2}{R^2}\right)^{n+1}\right\} - n^2 r^2\left(1 - \frac{r^2}{R^2}\right)\right],
\end{eqnarray}
\begin{eqnarray}
\nonumber
D(r) &=& -{p_0}{R^6} \left(1 - \frac{r^2}{R^2}\right) \\ \nonumber
&&\left[1 - \left\{(n + 4) \frac{r^2}{R^2} + (n - 1)\frac{r^4}{R^4}\right\}\right],
\end{eqnarray}
\[E(r) = \frac{1 + (n + 2)\frac{r^2}{R^2} - (2 n^2 + n - 1)\frac{r^4}{R^4}}{\left(1 + \frac{r^2}{R^2}\right)^{n + 2}}-1.\]

Thus, our model has four unknown parameters namely, $C$, $p_0$, $R$ and $n$ which can be fixed by the appropriate boundary conditions as will be discussed the following sections.

\section{\label{sec3}Boundary conditions}
At the boundary of the star $r=R$, we match the interior metric (\ref{d}) with the Schwarzschild exterior 
\begin{eqnarray}
\nonumber
ds^2 &=& \left(1 - \frac{2M}{r}\right) dt^2 - \left(1 - \frac{2M}{r}\right)^{-1}dr^2 - \\
&& r^2 (d\theta^2 + sin^2 \theta d\phi^2), \label{z}
\end{eqnarray}
which yields 
\begin{eqnarray}
R &=& \frac{2^{n+1} M}{2^{n} - 1},\label{aa}\\
C &=& 2^{-\left(n + p_0\right)} \\ \nonumber
&& ~exp\left[{\frac{p_0}{2} - \int\limits_{0}^{R} \left[\left(1 + \frac{r^2}{R^2}\right)^n-1\right]\frac{1}{r} dr}\right],\label{bb}
\end{eqnarray}
where $ M = m(R)$ denotes the total mass enclosed within a radius $R$. Eq.~(\ref{aa}) clearly shows that the compactness of the stellar configuration
$M/R$ will depend on the parameter $n$ which was not the case in the model previously developed by \cite{Sharma13}.

\section{\label{sec4}Bounds on the model parameters} 
For a physically acceptable stellar model, the following conditions should be satisfied:
\begin{itemize}
\item (i) $ \rho (r) \geq 0, ~~~ p_r (r) \geq 0, ~~~ p_{\perp}(r) \geq 0 $; 
\item (ii) $ \rho (r) - p_r (r) - 2p_{\perp} (r) \geq 0 $;
\item (iii) $ \frac{d\rho (r)}{dr} < 0, ~~~ \frac{dp_r (r)}{dr} < 0, ~~~ \frac{dp_{\perp} (r)}{dr} < 0 $;
\item (iv) $ 0 \leq \frac{dp_r}{d\rho} \leq 1 $, ~~~ $ 0 \leq \frac{dp_{\perp}}{d\rho} \leq 1 $.
\end{itemize}
Due to mathematical complexity, it is difficult to show analytically that our model complies with all the above mentioned conditions. However, by adopting numerical procedures, we have shown that for a specified bound all the above requirements can be fulfilled in this model.

Now, to get an estimate on the bounds of the model parameters, we note that $ p_r,~p_\perp \geq 0 $ at $ r = R $ if we have
\begin{equation}
p_0 \leq \frac{(2^n - 1)(2^n - 1 + n)}{2}.\label{ee}
\end{equation}
The strong energy condition $ \rho - p_r - 2 p_\perp \geq 0 $ at $ r = R $ puts a further constraint on the parameter $p_0$ given by 
\begin{equation}
p_0 \geq \frac{3(1-n)}{2} + (n - 4 + 2^n)2^{n-1}.
\label{ff}
\end{equation}
The condition $\frac{dp_\perp}{dr}\mid_{r = R} < 0 $  imposes the following constraint on $p_0$
\begin{equation}
p_0 > \frac{\left(n^2 - 2 \left(2^n-1\right)^2+2^n n \left(2^n-1\right)\right)}{\left(2 - 3n + 2^{n+1}\right)}.
\label{gg}
\end{equation}
The requirement $ \frac{dp_\perp}{d\rho}\mid_{r = R} < 1 $ puts the following bound
\begin{equation}
p_0 < 2^{n + 1} + n^2 - 2.
\label{jj}
\end{equation}
Similarly, the conditions $\frac{dp_\perp}{d\rho}\mid_{r = 0} < 1 $ and $\frac{dp_\perp}{d\rho}\mid_{r = R} < 1 $, respectively puts the following constraints on $ p_0 $:
\begin{equation}
p_0 < 8 + 2n - \sqrt{64 + 22n - 9n^2},
\label{kk}
\end{equation}
\begin{equation}
p_0 < \frac{4(2^{n+1}+n^2 - 2) - 2(2^n - 1)^2 + 2^n (2^n - 1) + n^2}{2^{n + 1} - 3n + 2}. 
\label{ll}
\end{equation}
All the above constraints when put together provides an effective bound
\begin{equation}
\frac{n^2 - 2 \left(2^n-1\right)^2+2^n n \left(2^n-1\right)}{2 - 3n + 2^{n+1}} < p_0 \leq \frac{(2^n - 1)(2^n - 1 + n)}{2}
\label{mm}
\end{equation}
on $p_0$ and $n$.

\subsection{Stability}
\begin{figure}[h]
\plotone{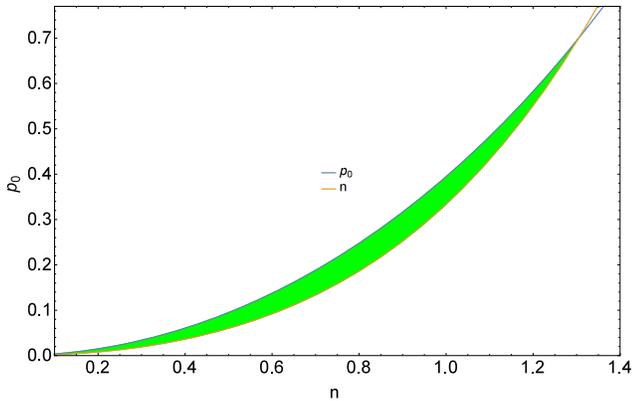}
\caption{Bounds on the model parameters $ p_0 $ and $ n $ based on physical requirements and stability. \label{fig1}}
\end{figure}

Though we have obtained an effective bound on $p_0$ and $n$ based on requirements (i)-(iv), a more stringent bound on these parameters may be obtained by analyzing the stability of the system. To check stability, we have followed the method of \cite{Herrera92} which states that for a potential stable configuration we should have $ {(\upsilon^2_{\perp} - \upsilon^2_{r})}\mid_{r = 0} < 0 $. In our case, the difference between the radial speed of sound $\upsilon^2_{r} (= \frac{dp_r}{d\rho})$ and  tangential speed of sound $\upsilon^2_{\perp} (= \frac{dp_{\perp}}{d\rho})$ evaluated at the centre $r=0$ is obtained as 
\begin{equation}
(\upsilon^2_{\perp} - \upsilon^2_{r})\mid_{r = 0} = -\frac{3 n^2+\left(p_0-8\right) p_0}{10 n (n+1)}.
\label{nn}
\end{equation}
Then Herrera's stability condition implies
\begin{equation} 
p_0 < 4 - \sqrt{16 - 3n^2}.
\label{oo}
\end{equation}
Similarly, $ {(\upsilon^2_{\perp} - \upsilon^2_{r})}\mid_{r = R} < 0 $ yields
\begin{equation}
p_0 < n^2 - 2^n {(n - 4)} + 4^n {(n - 2)} - 2.
\label{pp}
\end{equation}
Combining (\ref{mm}), (\ref{oo}) and (\ref{pp}), the most appropriate bound on the model parameters is finally obtained in the form
\begin{equation}
\frac{n^2 - 2 \left(2^n-1\right)^2+2^n n \left(2^n-1\right)}{2 - 3n + 2^{n+1}} < p_0 < 4 - \sqrt{16 - 3n^2}.
\label{qq}
\end{equation}
It is to be noted that for a real valued upper bound on $ p_0 $ we must have $ n \leq \frac{4}{\sqrt{3}} $. In Fig.~\ref{fig1}, we have shown the possible range of $p_0$ and $n$ (shaded region) for which a physically acceptable stable stellar configuration is possible.  

\section{\label{sec5}Physical analysis}
Having derived a physically plausible model, let us now analyze the implications of the modified \cite{Finch89} ansatz. Note that in our description, two of the four unknown parameters can be determined from the boundary conditions (\ref{aa}) and (\ref{bb}) provided the mass is known. Since the condition $p_r(R) = 0$ is automatically satisfied, it provides no additional information about the unknowns. Therefore, $n$ and $p_0$ remain free parameters in our construction. For a chosen value of $n$, the parameter $p_0$ can be appropriately fixed from within the bound provided in (\ref{qq}). Thus, all the physically interesting quantities of the model can be evaluated if the mass $M$ is supplied. 

To examine the nature of physical quantities, we have considered the pulsar $4U 1820-30$ whose estimated mass and radius are given by $M = 1.58~M_{\odot}$ and $R = 9.1~$km, respectively \cite{Guver10a}. Assuming $M = 1.58~M_{\odot}$, we note that if we set the dimensionless parameter $n = 0.6154$ and $p_0 = 0.1211~$Mev~fm$^{-3}$, we get exactly the same radius as estimated by \cite{Guver10a}. Moreover, the compactness of the star can be made as high as $\sim 0.4543$ for an upper limit of $n \sim 1.38$. Similarly, we have considered some other well studied pulsars like PSR J1903+327 \citep{Freire11}, 4U 1608-52\citep{Rawls11}, Vela X-1\citep{Rawls11}, PSR J1614-2230\citep{Demorest10}, SAX J1808.4-3658\citep{Elebert09} and Her X-1\citep{Abubekerov08} and shown that the estimated masses and radii of these stars can also be obtained by making necessary adjustments in the values of $n$. In Table \ref{tbl-1}, we have given the appropriate values of the adjustable \emph{compactness parameter} $n$ for which one can obtain the predicted masses and radii of the stars considered here. Respective central density ($ \rho_0 $), surface density ($ \rho_R $), central pressure ( $ p_{r0} $ ) and compactness ($u = \frac{M}{R}$) have also been shown in the table. The difference in the values of these parameters for different choices of $n$ has also been shown. 

\begin{figure}
\plotone{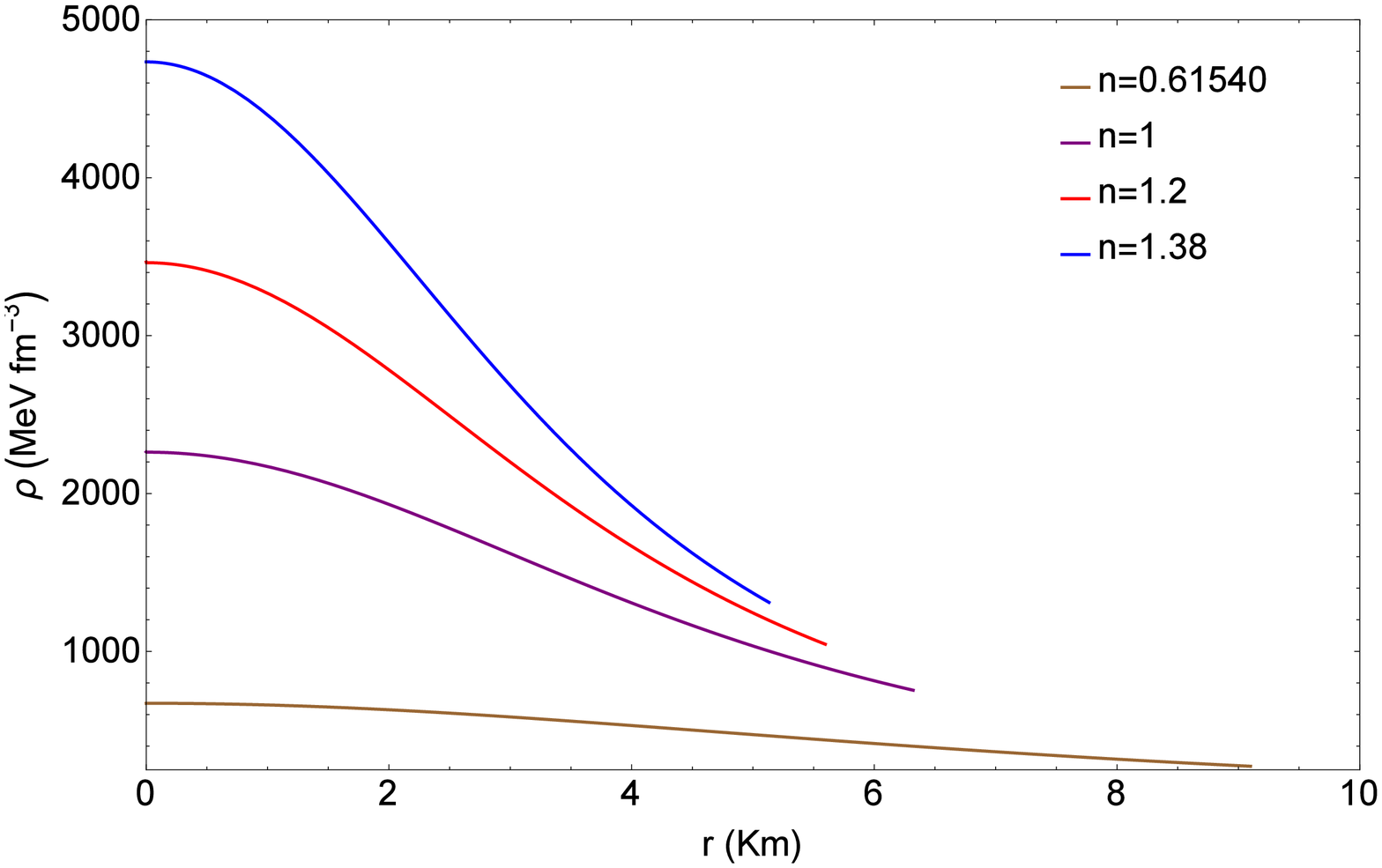}
\caption{Variation of density ($\rho$) against the radial parameter $ r $. \label{fig2}}
\end{figure}

\begin{figure}
\plotone{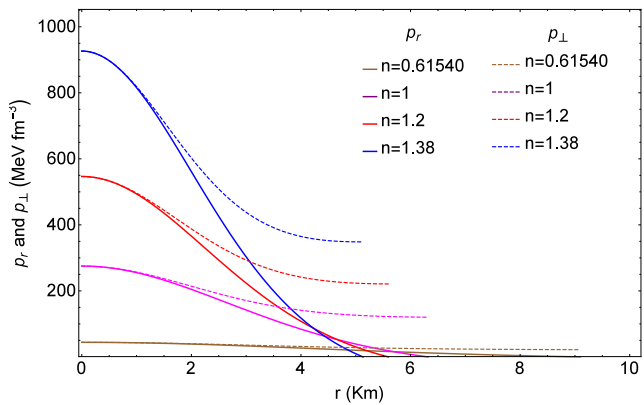}
\caption{Variation of radial ($ p_r $) and transverse ($ p_\perp $) pressure against the radial parameter $r$. \label{fig3}}
\end{figure}

\begin{figure}
\plotone{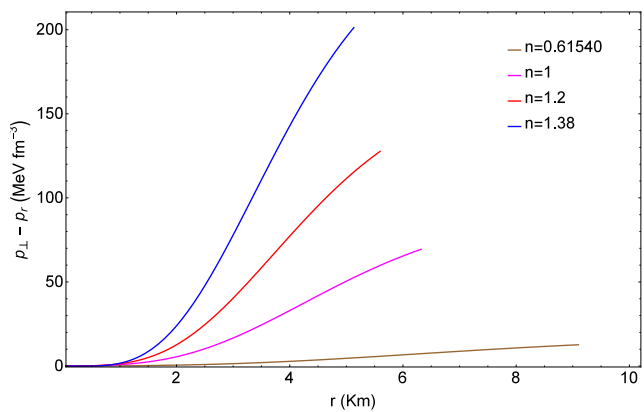}
\caption{Variation of anisotropy ($ p_\perp - p_r $) against  $r$. \label{fig4}}
\end{figure}

\begin{figure}
\plotone{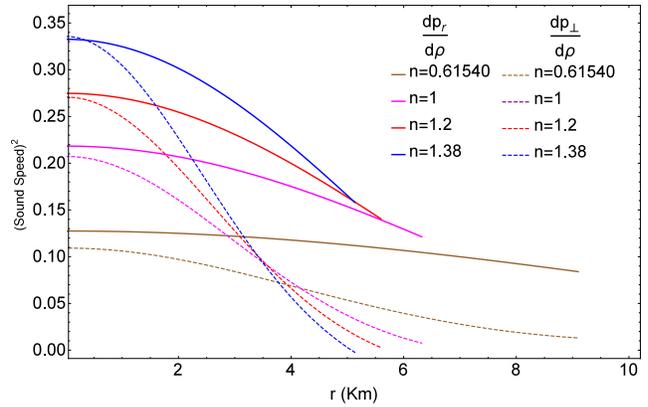}
\caption{Variation of $\frac{dp_r}{d\rho} $ and $\frac{dp_\perp}{d\rho} $ against the radial parameter $ r $.\label{fig5}}
\end{figure}

\begin{figure}
\plotone{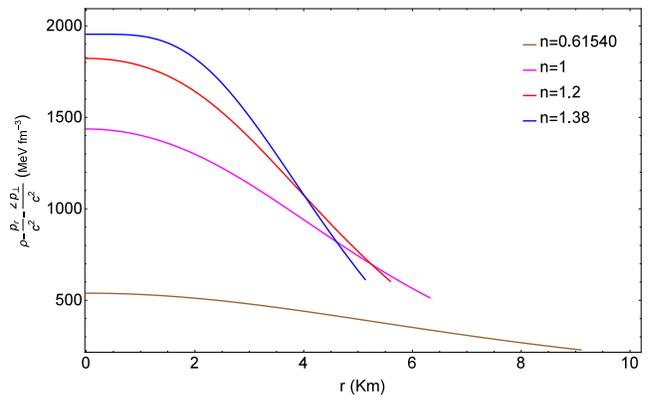}
\caption{($ \rho - p_r - 2p_\perp $) plotted against the radial parameter $ r $.\label{fig6}}
\end{figure}

\begin{figure}
\plotone{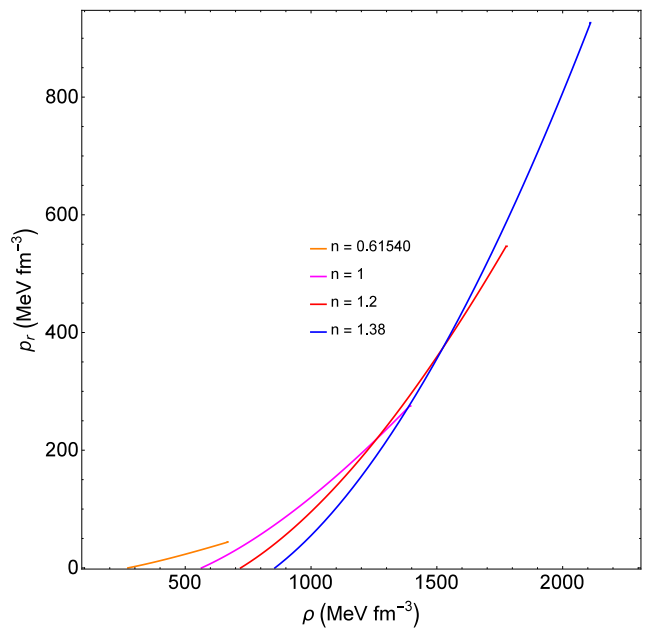}
\caption{Variation of radial pressure ($ p_r $) against density ($ \rho $) \label{fig7}}
\end{figure}

For a particular mass $M = 1.58~M_{\odot}$, we have also shown that all the physical quantities are well behaved at all interior points of the star within the specified bounds on $n$ and $p_0$. In Fig.~\ref{fig2}, we have shown the variation of density which shows that the density decreases from its maximum value at the centre towards the boundary. Moreover, the central density increases if the value of $n$ increases. In Fig.~\ref{fig3}, radial variation of the two pressures has been shown. As expected, the radial pressure $p_r$ vanishes at the boundary; however the tangential pressure $p_\perp$ remains finite at the boundary. As in the case of density, both pressures increase as $ n $ increases. In Fig.~\ref{fig4}, radial variation of anisotropy has been shown which shows that anisotropy is zero at the centre and is maximum at the surface. In Fig. \ref{fig5}, radial variations of sound speed in the radial and transverse directions have been shown which confirms that the causality condition is not violated throughout the configuration. In  Fig.~\ref{fig6}, we have plotted ($\rho - p_r - 2p_\perp$) which was shown to remain positive thereby implying that the strong energy condition is not violated in this model. Though we have not assumed any explicit EOS in our model, Fig.~\ref{fig7} shows how the radial pressure varies against the density for different values of $n$.\\ 

\begin{landscape}
\begin{deluxetable}{ccccccccc}
\tabletypesize{\small}
\tablecolumns{9}
\tablewidth{0pt}
\tablecaption{Estimation of physical values based on observational data. \label{tbl-1}}
\tablehead{
\colhead{\textbf{STAR}} & \colhead{\textbf{$ n $}} & \colhead{\textbf{$ p_{r0} $}}  & \colhead{\textbf{$ M $}} & \colhead{\textbf{$ R $}} & \colhead{\textbf{$ \rho_c $}} & \colhead{\textbf{$ \rho_R $}} & \colhead{\textbf{$ u (=\frac{M}{R}) $}} & \colhead{$\left(\frac{dp_r}{d \rho}\right)_{r=0} $} \\
 &  & MeV fm{$^{-3}$} & $ M_\odot $ & (Km) & MeV fm{$^{-3}$} & MeV fm{$^{-3}$} & & 
}
\startdata
\textbf{4U 1820-30} & \textbf{0.6154} & \textbf{0.1211} & \textbf{1.58} & \textbf{9.1} & \textbf{671.36} & \textbf{272.35} & \textbf{0.2561} & \textbf{0.1275} \\
  & 1    & 0.3638 &   & 6.32 & 2261.77 & 753.93  & 0.3688 & 0.2183  \\
  & 1.2  & 0.5667 &  & 5.59 & 3469.29 & 1047.59 & 0.4169 & 0.2748 \\
  & 1.38 & 0.8079 &   & 5.13 & 4737.27 & 1311.33 & 0.4543 & 0.3326 \\ \hline

\textbf{PSR J1903+327} & \textbf{0.6287} & \textbf{0.1269} & \textbf{1.667} & \textbf{9.438} & \textbf{637.65} & \textbf{256.89} & \textbf{0.2605} & \textbf{0.1303} \\
  & 1    & 0.3638 &   & 6.66 & 2036.74 & 678.91  & 0.3692 & 0.2183  \\
  & 1.2  & 0.5667 &   & 5.90 & 3114.30 & 940.39  & 0.4168 & 0.2748  \\
  & 1.38 & 0.8079 &   & 5.41 & 4259.59 & 1179.11 & 0.4545 & 0.3326  \\ \hline

\textbf{4U 1608-52} & \textbf{0.6752} & \textbf{0.1485} & \textbf{1.74} & \textbf{9.31} & \textbf{703.83} & \textbf{276.78} & \textbf{0.2757} & \textbf{0.1405} \\
  & 1    & 0.3638 &   & 6.96 & 1864.94 & 621.65  & 0.3688 & 0.2183 \\
  & 1.2  & 0.5667 &   & 6.16 & 2856.95 & 862.69  & 0.4166 & 0.2748 \\
  & 1.38 & 0.8079 &   & 5.65 & 3905.40 & 1081.06 & 0.4542 & 0.3326 \\ \hline

\textbf{Vela X-1} & \textbf{0.6672} & \textbf{0.1446} & \textbf{1.77} & \textbf{9.56} & \textbf{659.553} & \textbf{260.45} & \textbf{0.2731} & \textbf{0.1387} \\
  & 1    & 0.3638 &   & 7.08 & 1802.26 & 600.75  & 0.3688 & 0.2183  \\
  & 1.2  & 0.5667 &   & 6.27 & 2757.59 & 832.68  & 0.4164 & 0.2748 \\
  & 1.38 & 0.8079 &   & 5.75 & 3770.74 & 1043.79 & 0.4540 & 0.3326 \\ \hline

\textbf{PSR J1614-2230} & \textbf{0.7529} & \textbf{0.1892} &  \textbf{1.97} & \textbf{9.69} & \textbf{724.42} & \textbf{273.692} & \textbf{0.2999} & \textbf{0.1578} \\

  & 1    & 0.3638 &   & 7.88 & 1454.89 & 484.96  & 0.3688 & 0.2183  \\
  & 1.2  & 0.5667 &   & 6.98 & 2225.12 & 671.90  & 0.4163 & 0.2748 \\
  & 1.38 & 0.8079 &   & 6.40 & 3043.71 & 842.54  & 0.4540 & 0.3326 \\ \hline

\textbf{SAX J1808.4-3658} & \textbf{0.3703} & \textbf{0.0411} & \textbf{0.9} & \textbf{7.951} & \textbf{529.18} & \textbf{244.30} & \textbf{0.1669} & \textbf{0.0768} \\

  & 1    & 0.3638 &   & 3.6  & 6970.73  & 2323.58  & 0.3688 & 0.2183 \\
  & 1.2  & 0.5667 &   & 3.19 & 10653.30 & 3216.86  & 0.4161 & 0.2748 \\
  & 1.38 & 0.8079 &   & 2.92 & 14621.70 & 4047.46  & 0.4546 & 0.3326 \\ \hline

\textbf{Her X-1} & \textbf{0.3399} & \textbf{0.0344} &  \textbf{0.85} & \textbf{8.1} & \textbf{467.95} & \textbf{219.575} & \textbf{0.1548} & \textbf{0.1031} \\

  & 1    & 0.3638 &   & 3.40 & 7814.94  & 2604.98  & 0.3688 & 0.2183 \\
  & 1.2  & 0.5667 &   & 3.01 & 11965.50  & 3613.11 & 0.4165 & 0.2748 \\
  & 1.38 & 0.8079 &   & 2.76 & 16366.1 & 4530.33 & 0.4543 & 0.3326 \\
  
\enddata 
\end{deluxetable}
\end{landscape}

\section{\label{sec6} Discussion}
In this paper, we have solved the Einstein's field equations describing a spherically symmetric anisotropic matter composition by assuming the form of one of the metric potentials of the associated space-time and also by choosing a particular radial pressure profile. The assumed form of the metric potential is a generalization of the \citet{Finch89} anzatz, which has so far been utilized successfully by many authors to generate solutions to the Einstein's field equations in different astrophysical contexts. We note that a modification of the \cite{Finch89} ansatz for the metric potential $g_{rr}$ allows us to fit the theoretically obtained compactness to the observed compactness of a given star. We have shown that in the presence of such an adjustable parameter, it is possible to accommodate a large class of observed pulsars in our model. Another interesting feature of our approach is that though no a priori knowledge of the EOS is required in our set up, we have been able to show that the predicted masses and radii of the pulsars based on the exotic strange matter EOS formulated by \cite{Dey98} and examined by \citet{Gangopadhyay13} can also be fitted into our model.

\section*{Acknowledgements}

\noindent DMP is obliged for the support from the Inter-University Centre for Astronomy and Astrophysics (IUCAA), Pune, India, where a part of this work was carried out. RS acknowledges support from the IUCAA, under its Visiting Research Associateship Programme. DMP and VOT thank B S Ratanpal for useful suggestions.

\pagebreak

\end{document}